\documentstyle[11pt,newpasp,twoside,epsf]{article}
\def\edcomment#1{\iffalse\marginpar{\raggedright\sl#1\/}\else\relax\fi}
\reversemarginpar

\begin{document}
\title{On the Formation of Binary Stars and Small Stellar Groups
in Magnetically Subcritical Clouds}
 \author{Fumitaka Nakamura}
\affil{Faculty of Education and Human Sciences, Niigata University and
Astronomy Department, University of California at Berkeley}
\author{Zhi-Yun Li}
\affil{Department of Astronomy, University of Virginia, P.O. Box 3818,
Charlottesville, VA 22903}

\begin{abstract}
In the standard scenario of isolated low-mass star formation,
strongly magnetized molecular clouds are envisioned to condense
gradually into cores, driven by ambipolar diffusion. Once the
cores become magnetically supercritical, they collapse to form
stars.
Most previous studies based on this scenario are limited to axisymmetric
 calculations leading to single supercritical core formation.
The assumption of axisymmetry has precluded a detailed investigation of
 cloud fragmentation, generally thought to be a necessary step in the
 formation of binary and multiple stars.
In this contribution, we describe the non-axisymmetric evolution of
initially magnetically subcritical clouds using a newly-developed MHD
code. It is shown that non-axisymmetric perturbations of modest
fractional amplitude ($\sim 5\%$) can grow nonlinearly in such
clouds during the supercritical phase of
cloud evolution, leading to the production of either a highly elongated bar
or a set of multiple dense cores.
\end{abstract}

\section{Introduction}

Over the last few decades, a basic framework has been developed for the
formation of low-mass stars in relative isolation (Shu, Adams, \&
Lizano 1987). In this by now ``standard'' picture, a molecular cloud,
which is initially supported by strong magnetic field against its
self-gravity, gradually contracts as the magnetic support weakens
by ambipolar diffusion. Magnetically supercritical cores are
formed, which collapse to produce stars.  Quantitative studies based
on this scenario have been carried out by many authors. In most of
such studies, axisymmetry has been adopted.  However, observations
have shown that binary and multiple stars are common product of star
formation. We need to understand how such (non-axisymmetric) stellar
systems are formed in magnetically supported clouds. To elucidate the
formation mechanism of binary stars and stellar groups, we have begun
a systematic numerical study of the non-axisymmetric evolution
initially magnetically subcritical clouds, by removing the restriction
of axisymmetry. In this contribution, we present some of our recent
results on this investigation.

\section{Model and Numerical Method}

As a first step, we adopted the thin-disk approximation often used
in axisymmetric calculations (e.g., Basu \& Mouschovias 1994;
Li 2001). The disk is assumed in hydrostatic equilibrium in
the vertical direction. The vertically-integrated MHD
equations are solved numerically for the cloud evolution in the disk
plane, with a 2D MHD code (see Li \& Nakamura 2002 for code description).
The magnetic structure is solved in 3D space.

The initial conditions for star formation are not well determined either
observationally and theoretically.  Following Basu \& Mouschovias
(1994), we prescribe an axisymmetric reference state.
See Nakamura \& Li and Li \& Nakamura (2002) for the details of the
reference cloud model.
The reference cloud is allowed to evolve into an equilibrium
configuration, with the magnetic field frozen-in.  Once the equilibrium
state is obtained, we reset the time to $t=0$ and
add a non-axisymmetric perturbation to the surface density distribution.
Then, the cloud evolution is followed with the ambipolar diffusion turned
on.

\section{Numerical Results}

From axisymmetric calculations, Li (2001) classified the evolution of
magnetically subcritical clouds into two cases,
depending mainly on the initial cloud mass and the initial
density distribution.
When the initial cloud is not so massive and/or has a
centrally-condensed density distribution,
it collapses to form a single supercritical core
({core-forming cloud}).
On the other hand, when the initial cloud has
many thermal Jeans masses and/or a relatively flat density distribution
near the center, it collapses to form a ring after the central region
becomes magnetically supercritical
({ring-forming cloud}).
In the following, we show that
the core-forming cloud doesn't fragment
during the dynamic collapse phase, but becomes unstable to the bar mode
({\it bar growth}), whereas the ring-forming cloud
can break up into several blobs
({\it multiple fragmentation}).

\subsection{Bar growth: Implication for Binary Formation}

In Fig.~1 we show an example of the bar growth models. In this model, we
adopted the reference density distribution of Basu \& Mouschovias (1994),
which is more centrally-condensed than the model to be shown in the next
subsection, and the rotation profile of Nakamura \& Hawana (1997).
It has a
characteristic radius of $r_0=7.5\pi c_s^2/(2\pi G\Sigma_{0,\rm ref})$
(where $c_s$ is the effective isothermal sound speed and $\Sigma_{0,
\rm ref}$ the central cloud surface density in the reference state),
initial flux-to-mass ratio of
$\Gamma _0 = 1.5 B_{\infty}/(2\pi G^{1/2}\Sigma_{0,\rm ref})$
(where $B_\infty$ is the strength of the initially uniform
background field), and
a dimensionless rotation rate of $\omega=0.1$.
We added to the equilibrium state an $m=2$ perturbation of surface
density, with a fractional amplitude of merely 5\%.
During the initial quasi-static contraction phase, a central core
condenses gradually out of the magnetically subcritical cloud, with no
apparent tendency for the mode to grow. Rather, the iso-density contours
appear to oscillate, changing the direction of elongation along $x$-axis
in the disk plane to $y$-axis.  After a supercritical core develops, the
contraction becomes dynamic and the bar mode grows significantly.
During the intermediate stages [panels (c) and (d)],
the aspect ratio of the bar
remains more or less frozen at $R\sim 2$.  As the collapse continues,
the growth rate of the bar increases dramatically by the very end of
the starless collapse.
The density distribution along the minor axis of the bar is well
reproduced by a power-law profile of $r^{-2}$, which is different from
that of an isothermal equilibrium filament ($\propto r^{-4}$).
When the volume density exceeds a critical value of $10^{12}$ cm$^{-3}$,
we changed the equation of state from isothermal to adiabatic, to mimic
the transition to the optically thick regime.
The bar is surrounded by an accretion shock, which is analogous to the
first core of spherical calculations [panel (f)].
The aspect ratio of this ``first''
bar continues to increase during the early optically thick regime.
The highly elongated first bar is expected to break up into two or more
pieces.  We suspect that bar fragmentation is an important, perhaps
the dominant, route for binary and small multiple-star formation.

\begin{figure}
\plotone{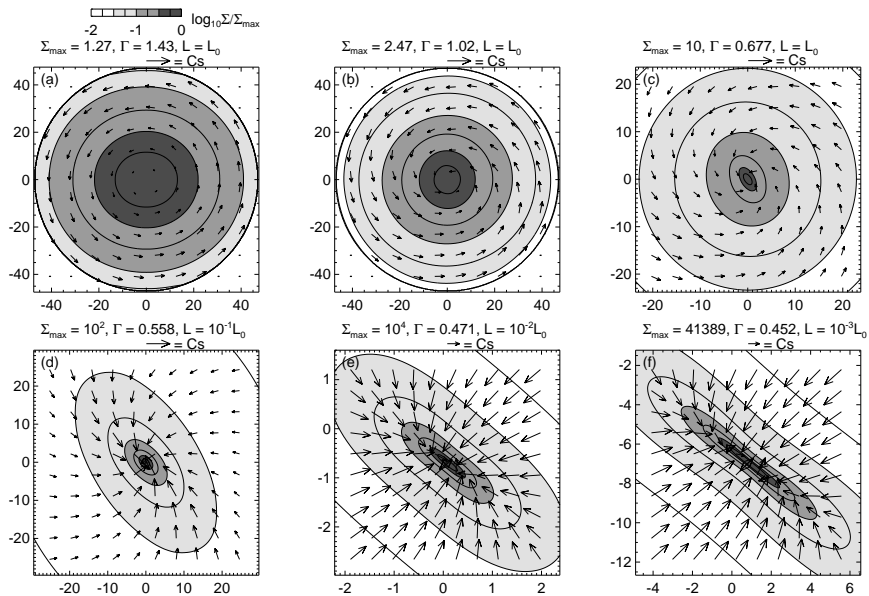}
\caption{\small {\it Bar Growth}: Time evolution of surface density and
velocity distributions for the
model with $(\Gamma_0, r_0, \omega)= (1.5, \, 7.5\pi, \, 0.1)$.
[(a) t=0, (b) 17.49, (c) 24.46, (d) 25.21, (e)
25.2500, and (f) 25.2503].
We added to the equilibrium cloud 
an $m=2$ density perturbation with a fractional amplitude of 5\%.
If we choose $\Sigma_{0,\rm ref} = 0.01$ g cm$^{-3}$,
$T_{\rm eff} = 30$ K, then the units for time, length and speed would
be $t_0 = 2.46\times 10^5$ yr, $L_0=0.082$ pc and $c_s=0.33$~km~s$^{-1}$.
The maximum volume densities in panels (a) through (f)
would be (a) $4.46\times 10^3$
cm$^{-3}$, (b) $1.70\times 10^4$ cm$^{-3}$, (c)
$2.79\times 10^5$ cm$^{-3}$ (d) $2.80\times 10^7$ cm$^{-3}$,
(e) $2.78\times 10^{11}$  cm$^{-3}$, and (f) $4.77\times 10^{12}$
cm$^{-3}$, respectively.
}
\end{figure}

We have also followed the evolution of this model cloud perturbed by
other (higher) $m$ modes ($m\ge3$), and found no significant mode
growth. The reason why the cloud is unstable only to the bar mode
appears to be the following. In the absence of nonaxisymmetric
perturbations, the supercritical collapse approaches a self-similar
solution derived approximately by Nakamura \& Hanawa (1997). In the
self-similar solution, the effective radius of the central
plateau is at most 3-4 times the effective Jeans length, making
the cloud unstable to dynamic contraction but not to multiple
fragmentation. Indeed, Nakamura \& Hanawa (1997) showed that the
self-similar solution is unstable only to the $m=2$ mode, consistent
with our result.  The tendency for the supercritical collapse
to approach the self-similar solution is responsible for the bar
formation during the dynamic collapse. Detailed numerical results on bar
formation will appear elsewhere (Nakamura \& Li 2002, in preparation).

\subsection{Multiple Fragmentation and Formation of Small Stellar Groups}

\begin{figure}
\plotone{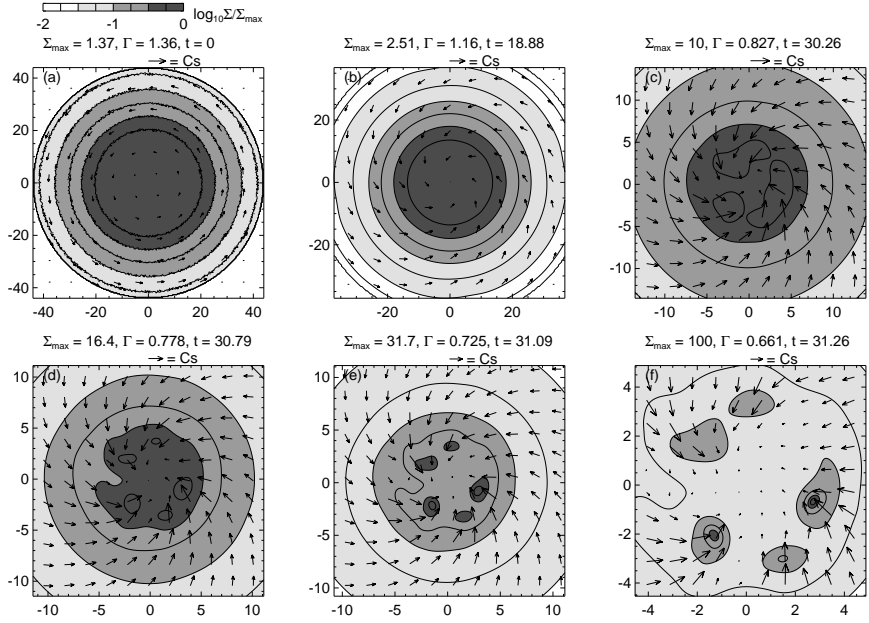}
\caption{\small {\it Multiple Fragmentation}:
Same as Figure 1 but for the 
model with $(\Gamma_0, r_0, \omega)= (1.5, \, 7\pi, \, 0.1)$.
We added to the equilibrium cloud 
random density perturbations with a maximum fractional amplitude of 10\%.
If we choose $\Sigma_{0,\rm ref} = 0.01$ g cm$^{-3}$,
$T_{\rm eff} = 30$ K, then the units for time, length and speed would
be $t_0 = 2.46\times 10^5$ yr, $L_0=0.082$ pc and $c_s=0.33$~km~s$^{-1}$.
The maximum volume densities in panels (a) through (f)
would be (a) $5.19\times 10^3$
cm$^{-3}$, (b) $1.86\times 10^4$ cm$^{-3}$, (c)
$2.77\times 10^5$ cm$^{-3}$ (d) $9.39\times 10^5$  cm$^{-3}$,
(e) $3.51\times 10^{6}$  cm$^{-3}$, and (f) $2.77\times 10^{7}$
cm$^{-3}$, respectively.
}
\end{figure}
In Fig.~2 we show an example of the multiple fragmentation models.
In this model, we adopted the reference density profile of Li (2001)
with $n=8$,
which is less centrally-condensed than the model shown in the previous
subsection.
The model has a characteristic radius of
$r_0=7\pi c_s^2/(2\pi G\Sigma_{0,\rm ref})$,
initial flux-to-mass ratio of
$\Gamma _0 = 1.5B_{\infty}/(2\pi G^{1/2}\Sigma_{0,\rm ref})$, and
rotation rate of $\omega=0.1$.
Random density perturbations are added to the axisymmetric equilibrium state.
The maximum fractional amplitude of the perturbations is set to
10\%.
During the quasi-static contraction phase, the infall motions
are subsonic, and there is no sign of fragmentation.
Once the flux-to-mass ratio in the central high-density region drops
below the critical value, the contraction is accelerated near the center.
As the collapse continues, the central supercritical region
begins to fragment into five blobs.
By the time shown in panel (f), the blobs are well separated from
the background material and are significantly elongated.
Subsequent dynamic collapse of each blob is similar
to that of the bar growth case. Individually, we expect each core
to produce a highly elongated bar, which could further break up
into pieces, producing perhaps binary or multiple stars. Together,
the formation of a small stellar group is the most likely outcome.
Detailed numerical results on multiple fragmentation are given in
Li \& Nakamura (2002).

\section{Summary}

Our main conclusion is that despite (indeed because of) the presence of
the strong magnetic
field, the initially magnetically subcritical clouds are unstable
to non-axisymmetric perturbations during the supercritical phase
of cloud evolution.
The cloud evolution is classified into two cases, depending mainly on
the initial cloud mass and density distribution.
When the initial cloud is not so massive and has a centrally condensed
density distribution,
it doesn't break into pieces but becomes unstable to a bar mode
({\it bar growth}).  This bar is expected to fragment into two or
more pieces to form binary or small multiple stars,
when the bar becomes opaque to dust emission and is
surrounded by an accretion shock.
On the other hand, when the initial cloud has many Jeans masses and
a relatively flat density distribution near the center,
it can fragment into several or many cores after a supercritical region
develops near the center ({\it multiple fragmentation}).
This fragmentation may be responsible for small cluster formation in
relatively isolated regions.

Boss (2000) showed the fragmentation of 3D magnetic clouds numerically,
treating the magnetic forces and ambipolar diffusion in an approximate
way (see also the contribution by Boss).
He concluded that  magnetic fields (magnetic tension force)
can enhance cloud fragmentation
by reducing the tendency for the development of a central singularity,
which would make fragmentation more difficult.  We also find that
magnetic fields can have beneficial effects on fragmentation.
Strong magnetic fields can support clouds with many Jeans masses and
flatten mass distribution, both of which are conducive to fragmentation
once the magnetic support weakens through ambipolar diffusion.

Numerical computations in this work were carried out
 at the Yukawa Institute Computer Facilities, Kyoto University.
F.N. gratefully acknowledges the support of the
JSPS Postdoctoral Fellowships for Research Abroad.

\end{document}